# Peer Instruction: Comparing Clickers to Flashcards

**N. Lasry**, John Abbott College, Montreal, Qc

*Peer Instruction* (PI) is a student-centered instructional approach developed at Harvard by Eric Mazur (1997). The method has been welcomed by the science community and adopted by a large number of colleges and universities, due among other reasons to its common sense approach and its documented effectiveness (Fagen *et al*, 2002; Crouch & Mazur, 2001, Mazur, 1997). In PI, the progression of any given class depends on the outcome of real-time student feedback to ConcepTests: multiple-choice conceptual questions. In the early 1990s, students responded to ConcepTests using flashcards showing their answer. Instructors would then estimate the proportion of students holding each alternative conception. A few years later Mazur began using wireless handheld devices - colloquially called 'clickers'- to replace the flashcards. Previous users of clickers in university classrooms had reported benefits such as increased rates of attendance and decreased rates of attrition (Owens *et al., 2004;* Lopez-Herrejon & Schulman, 2004). The purpose of this paper is to determine the specific contribution of 'clickers' to conceptual learning and traditional problem solving skills as compared to low-budget flashcards.

## *Study Description*

First-semester students in a two-year Canadian public community college were randomly assigned by the registrar to one of two sections of an algebra-based mechanics course. Instruction in the fist section consisted of PI with clickers (n=41) while the other followed PI with flashcards (n=42) to respond to in-class ConcepTests. Both sections were taught by the author, followed the same course structure and content (using 3-4 ConcepTests with peer discussion in each class) and had the same laboratory component. A schematic description of the PI method used in this study is shown below (Fig.1).

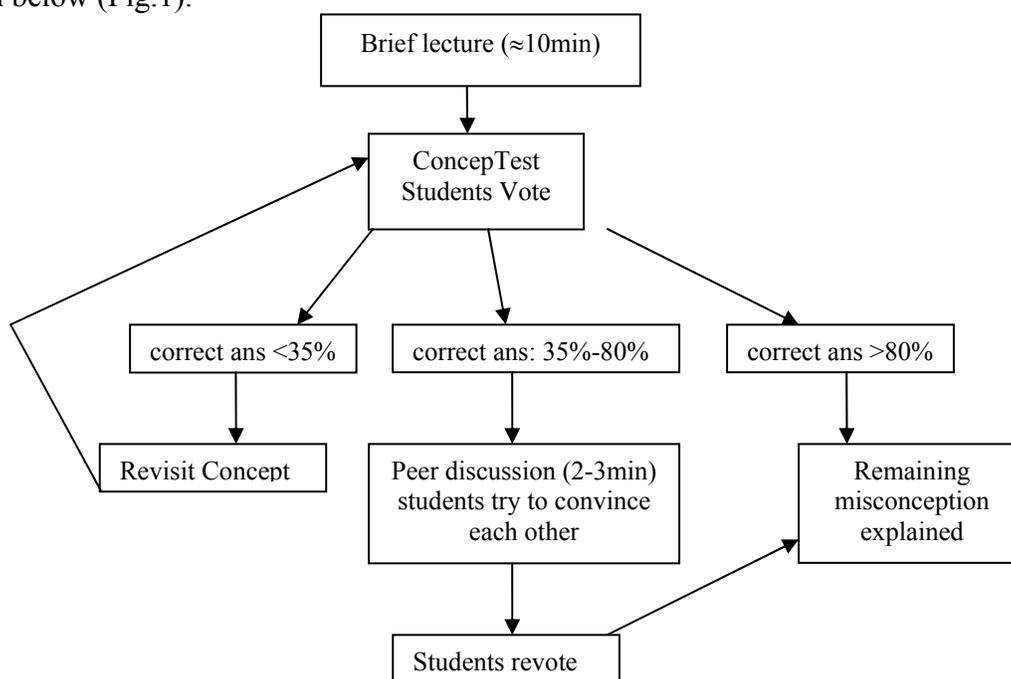

Figure 1  A *Peer Instruction* Implementation Algorithm

Conceptual understanding was measured in both groups the first and last week of the semester with the Force Concept Inventory (Halloun et al., 1995; Hestenes et al., 1992). Traditional problem solving skills were assessed using the college's comprehensive final examination. This exam was constructed by a committee of physics instructors (none of which were involved with this study) and had to be approved unanimously by all those teaching the course (10-12 instructors). Each instructor marked a single exam question for the entire cohort (n ≈500; not just for his or her students n ≈40). This insured that no group had an exam of a differing difficulty, or a corrector of different generosity. Furthermore, the correctors of the exam questions were unaware of which students belonged to which treatment condition.

## *Results: The effect of clickers on learning*

To determine the specific contribution of clickers on learning, the FCI pre-test, FCI post-test, FCI normalized gain and exam data are compared for both PI groups below:

**Table 1**
The effect of clickers: difference in learning data between flashcard and clicker groups

|  | PreFCI /30 | PostFCI /30 | g | Exam (%) |
|---|---|---|---|---|
| **Clickers (n= 35)** | 11.9 | 19.9 | 0.486 | 69.8 |
| **Flashcards (n=34)** | 13.6 | 21.3 | 0.520 | 71.6 |
| *t-test (2-tailed) p* | 0.209 | 0.351 | 0.745 | 0.630 |

These results show that both groups did not differ significantly in FCI score neither at the beginning of the semester (p=0.209) nor at its end (0.351). The use of clickers does not seem to add to the amount of conceptual learning or to problem solving skills. Indeed, although clickers have been reported to have a motivating influence, over the course of a semester no significant differences were found in conceptual learning gains (p = 0.745) nor in traditional problem solving skills (0.630). This implies that PI is an effective instructional approach which is independent from the use of technology such as clickers.

## **Lack of added effectiveness with clickers**

Against expectations, clickers do not provide any additional learning benefit to students. Previous users of clickers in university classrooms had reported benefits such as increased rates of attendance and decreased rates of attrition (Owens *et al., 2004;* Lopez-Herrejon & Schulman, 2004). However, no data was found in this study to support the claim that clickers increase conceptual learning. PI is an elaborate pedagogical approach that emphasizes basic concepts, has students commit to a conception, provides a setting for peer discussion and has instructors explicitly address misconceptions. Clearly, the technology is not the pedagogy. But if clickers don't add to learning should they be abandoned?

In fact, clickers should be greatly encouraged. Although this conclusion seems to contradict the previous finding, there are three main reasons why clicker use should continue to be encouraged.. First and foremost, clickers are responsible for much of the attention given to the PI approach. Indeed, much of the success of PI implicitly rest on the use of clickers (Burnstein & Lederman, 2003, 2001). Many instructors, including myself, have adopted the approach due to the appeal of using this technology in their classrooms. Using PI with clickers however forces instructors to reconsider their teaching, focus on concepts and thus fundamentally reshape their instruction. Since many instructors would not give proper attention to PI were it not for the clickers, one must continue to encourage their use.

Second, using clickers in the classroom allows instructors to archive ConcepTest data. Beyond data analyses and research questions that can be later addressed, this data can be used instructionally to sort out useful ConcepTests from those that work poorly. Furthermore, ConcepTests of questionable effectiveness could be reformulated and a core set of questions can evolve from one semester to another. Using flashcards does not enable the instructor to collect any ConcepTest related data. Thus, reusing the same questions from semester to semester may differ in effectiveness from using questions that can be modified from one semester to the next. Since only one semester of implementation was compared no such differences were found although differences may be expected to emerge over time.

The third reason for encouraging clicker use is to maximize the effect of peer discussions. Currently, 2-way clickers with LCD displays are available. These clickers allow students to send data but also receive data from the instructor's computer (such as acknowledgment of vote reception). To maximize the effect of peer discussions, one may program the response displayed to students so that it pairs students of differing conceptions. The response could then relocate a student to another seat in the classroom where the adjacent student holds a different conception. Using the clicker-display to pair students holding different conceptions would therefore have the potential to change the quality of peer-to-peer discussion and ultimately the effectiveness of PI in ways that are impossible with flashcards.

## *Conclusion*

Some instructors may be aware of PI methodology and willing to reshape their instruction to provide greater focus on basic concepts. Yet, the capital expense for the purchase of clickers and related hardware may not be available and passing the expense onto the students may not possible or desirable. In this instance, PI should be implemented with flashcards as it is the PI pedagogy which is effective regardless of the modality used by students to report their answer. Indeed, although clickers have many advantages, their use does not increase conceptual learning or traditional problem solving skills. Thus, one should focus on the instructional approach provided by PI and not conflate the technology with the pedagogy.

The author is grateful to Eric Mazur and Jessica Watkins for valuable comments and discussions on the manuscript. This study was supported by the Programme d'Aide sur la Recherche en Enseignement et en Apprentissage (PAREA) of the Ministère de l'Éducation du Québec